# Intro to Quantum Harmony:
# Chords in Superposition

Christopher Dobrian [1] and Omar Costa Hamido [2]

[1] University of California, Irvine CA 92697-2775, USA
`dobrian@uci.edu`

[2] Centre for Interdisciplinary Studies (CEIS20), University of Coimbra, Portugal
`ocostaha@uci.edu`

**Abstract.** Correlations between quantum theory and music theory—specifically between principles of quantum computing and musical harmony—can lead to new understandings and new methodologies for music theorists and composers. The quantum principle of *superposition* is shown to be closely related to different interpretations of musical meaning. Superposition is implemented directly in the authors' simulations of quantum computing, as applied in the decision-making processes of computer-generated music composition.

## 1      Quantum Theory and Music Theory

### 1.1      Superposition

The purpose of this article is to draw correlations between quantum theory and music theory, specifically between principles of quantum computing and musical harmony. We believe that these correlations may be interesting and useful to music theorists and composers. We will demonstrate how these relationships can form new metaphors for understanding musical harmony, and we will provide some possible concrete implementations in experimental arenas such as computer-generated composition.

The most notable correspondence is derived from the quantum principle of *superposition*.[1] In quantum computing, a fundamental unit known as the *qubit* (**qu**antum **b**inary dig**it**) does not have an established state—0 or 1—as a bit does in conventional computing. Rather, the qubit is in an unmeasurable state which can be thought of as the superposition (linear combination) of two *basis states* (vectors) known as |0⟩ and |1⟩ ("ket 0" and "ket 1"). In practice, one might think of a qubit as having a probability of being 0 and a probability of being 1, with the sum of those two probabilities being 1, but those probability values are only knowable by measuring (observing) the qubit, which has the effect of collapsing it to a classical representation of 0 or 1.

---

[1] C.f., [1, 2].

This quality of being unknowable until observed is known to most people through the thought experiment of Schrödinger's cat. In that hypothetical, a cat is placed for one hour in a box, in which is also placed a contraption that may or may not release poisonous gas into the box, depending on whether or not a single radioactive atom decays in that hour. During that hour, the alive-or-dead state of the cat is in superposition, unknowable until the moment of observation, when the box is opened. The thought experiment demonstrates that 1) a microcosmic happenstance on the order of a single atom (viz. a single qubit) can have a larger-scale effect, 2) the classical binary (alive or dead, 1 or 0) is reified only by observation, and 3) until that moment of resolution, the situation is ambiguous (except to the cat).

Music theorists and composers deal with a metaphorically comparable sort of ambiguity all the time. The meaning[2] of a given musical chord, in terms of its tonal function in its current context, frequently may have more than one interpretation. Music analysts strive either to remove that ambiguity by identifying the "true" function of the chord, or to draw some conclusions about the composer's process based on the way the composer treats the ambiguity.

The role of a composer is different from that of a theorist. One important distinction is between a theorist's analytical viewpoint, looking at existing music and trying to discern why and how it works, and a composer's creative viewpoint, looking at musical potential and trying to decide where to go. In the case of the analyst, the music pre-exists as a fixed entity, and the progress from one chord to another is a known and unalterable fact. In the case of a composer, by contrast, the musical "present" contains infinite potential to proceed anywhere, because the composer has free will. Infinite free will, however, is terrifying and unmanageable for most composers, which is why they self-impose constraints that limit their possibilities[3].

### 1.2     Chord Ambiguity and Equal Temperament

Composers exploit the inherent ambiguity of chords as a means to pivot to a more or less unexpected subsequent chord. This is a well-known theoretical concept, one which has enabled composers for centuries to achieve musical variety and modulate from one tonal center to another. Indeed, in jazz and in classical music of the last 150 years, the listener may be suspended in a state of constant ambiguity as to where the music is headed. Constant chromaticism, constant modulation, and atonality all utilize tonal ambiguity, and are all made possible by the inherent neutrality of twelve-tone equal temperament.

---

[2] C.f., [3].

[3] "I experience a sort of terror when, at the moment of setting to work and finding myself before the infinitude of possibilities that present themselves, I have the feeling that everything is permissible to me... Will I then have to lose myself in this abyss of freedom?... I have no use for a theoretic freedom. Let me have something finite, definite — matter that... presents itself to me together with limitations. I must in turn impose mine upon it."
— Igor Stravinsky [4], pp. 63-64

Twelve-tone equal temperament is the tuning system that is generally accepted as the norm in Western music, and that is assumed by every Western harmony text, be it classical, jazz, atonal, etc. Yet, tonal harmonic theory is derived from the diatonic scale, which grew out of just intonation and vastly predates equal temperament. Therefore, there is an inherent tension between the uniform intervals of the chromatic scale in equal temperament and the just intonation that is implicit in the diatonic scale and its chromatic alterations. This is common knowledge to most theorists, yet it is considered important in some contexts and conveniently ignored in others, a paradox made possible by the use of twelve-tone equal temperament.

For example, the fundamental tonal concept of the "circle of fifths" depends on the enharmonic equivalency inherent in equal temperament. A stack of twelve perfect fifths only creates a circle because, at some point on the circle, a pitch is respelled as its enharmonic equivalent, such as F♯/G♭ (figure 1).[4]

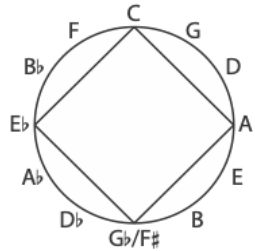

**Fig. 1.** A diminished seventh chord depicted on the circle of fifths.

Classical composers, theorists, and analysts quite reasonably insist on correct spelling of pitches consistent with the underlying tonal theory. For example, the third of a G♯ major triad is spelled B♯ rather than C. Theory textbooks and analyses distinguish between the interval of a diminished fourth and a major third, or between a minor seventh and an augmented sixth, in terms of their harmonic role, while in the same text also touting the interchangeability of enharmonic spellings for purposes of tonal ambiguity and modulation. The theory insists that these enharmonically spelled intervals, which are in fact identical in equal temperament, are also paradoxically different because of their tonal function, but that that difference can magically evaporate through mental respelling. Indeed, that "magic" is what enables much of the interesting and beautiful chromaticism of romantic music and jazz.

---

[4] In equal temperament, the interval of a perfect fifth is tempered by -1.955 cents ($^1/_{12}$ of the Pythagorean comma) relative to a just fifth, such that after ascending twelve perfect fifths from C, B♯ is exactly the same as C.

## 2      Superposition and the Diminished Seventh Chord

The most obvious example of this magic is the diminished seventh chord. The utility of the diminished seventh chord as a means of misdirection or modulation relies on enharmonic respelling, especially the enharmonic identity between the minor third and the augmented second. A diminished seventh chord is a stack of three minor thirds, causing the interval of an augmented second from the seventh of the chord going up to the root (figure 2).

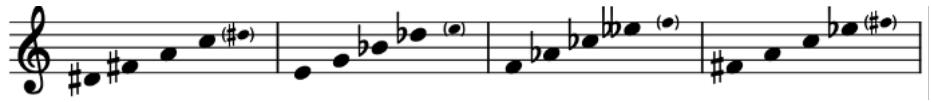

**Fig. 2.** The intervals of the diminished seventh chords.

The sound of the augmented second is identical to that of the minor third, so the diminished seventh chord effectively divides the circle of fifths equally into four minor thirds (figure 1). Any diminished seventh chord can be spelled in four different ways, by considering any of its four pitches to be the root of the chord (figure 3). By implication, each note of the chord can therefore potentially function in any of four ways[5], as either the root, third, fifth, or seventh of the chord. Similarly, the two tritones present in the diminished seventh chord are each ambiguous in their function due to these enharmonic reinterpretations, potentially functioning as the interval of either an augmented fourth or a diminished fifth.

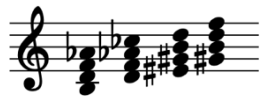

**Fig. 3.** Respellings of the same set of four pitch classes.

The essential function of the diminished seventh chord is as a leading tone chord in minor tonalities, or as an altered (borrowed) leading tone chord in major keys. In both instances, one chromatic alteration is needed, because the diminished seventh chord doesn't occur naturally in either the major scale or the natural minor scale. The root of the chord functions as a leading tone, wanting to resolve upward by semitone, and the diminished seventh of the chord inherently pulls downward by semitone. So, since any of the four pitches of the chord may function as the root, any time that a diminished seventh chord occurs, it has the potential to proceed to (at least!) eight different chords, four major chords and four minor chords. This ambiguity and versatility led to the diminished seventh chord's most common usage, as a secondary leading tone seventh chord (an implied secondary dominant), used either as a fleeting "tonicization" of a non-tonic degree of the scale without actually leaving the home key, or as an important force in modulating to a new key (figure 4).

---

[5] We will omit, for now, "wrong" spellings, such as B♯ E♭ F♯ B♭♭.

**Fig. 4.** Secondary diminished seventh chords as passing chords and as a modulating device.

The important thing to note for our discussion of this chord's relationship to quantum theory is that, for the listener, the chord's function—its directionality and its tonal meaning—is highly ambiguous until the moment of its resolution. One can say that the diminished seventh chord is a complex quantum state, and its resolution is the measurement of that quantum state. For a composer or an improviser, the chord is in a state of multiplicity (viz. superposition) until a decision is made to choose one of the multiple possible resolutions. For the listener, the chord may seem at first to have a primary "most likely" interpretation, but the chord's true function is evident only in retrospect, based on the chord that succeeds it. This is not to suggest that listeners are literally performing realtime functional analysis of the chords they hear, but that the ambiguity of the diminished seventh chord, and its multiple enharmonic representations, permit the composer or improviser to conceptually reinterpret (effectively, respell) the chord to have a different root, to lead to a different, less expected chord. The resolution, which may be unexpected, makes musical sense in light of that conceptual respelling.

In the example below (figure 5), starting in the key of C, the second chord is heard initially as a C♯°7 chord leading to ii. Once we land on the third chord, however, we retrospectively reinterpret (respell) the soprano note of the second chord from B♭ to A♯. From the third chord, B major, the bass note of the fourth chord only really makes sense as an E♯ leading to F♯. However, on the downbeat of the second measure it falls to E, revealing retrospectively that the E♯ may actually have been functioning as an F (or alternatively, that the A major chord may be some sort of deceptive cadence in F♯ minor). From there, in the second chord of the second measure, the tenor note can be heard either as B♯ leading us to C♯ or as C falling to B. The correct interpretation is revealed to us only retrospectively when we arrive on the E major chord.

**Fig. 5.** Modulation via respelled secondary diminished seventh chords.

Because of the diminished seventh chord's inherent ambiguity, when it is used consecutively or in sequence, or is reinterpreted in succession and points to far-related tonal centers, the listener tends to lose a sense of tonal gravity, and becomes less confident in the directionality of the music. This ability for the chord to create a sense of wandering, and to go almost anywhere, led composer and theorist Arnold Schoenberg to refer to it as a vagrant (*vagierende*) chord [5].

The foregoing discussion is known territory to music theorists, yet the relevancy to quantum computing has rarely[6] been noted. We have demonstrated that this particular chord shares the key characteristics of Schrödinger's cat: 1) a microcosmic occurrence (in this case, which note of the diminished seventh chord is considered the root) has a larger-scale effect (the progression to the subsequent chord and possibly to a far-removed key), 2) the true nature of the chord is reified only by observation of its resolution, and 3) until that moment of resolution, the situation is ambiguous; the chord is in superposition.

## 3   Practical Application: Quantum Decision Making

So much for the metaphorical correlations between harmonic ambiguity and quantum superposition. But what does this imply in terms of its practical applications? We will direct our attention first to the potential use of quantum circuits in computer-generated composition. The probabilistic nature of quantum computing makes it useful for conceptualizing and implementing computer decision making in tonal chord progressions. In this initial discussion, we continue using the well-studied diminished seventh chord as an archetypical example of an ambiguous chord, and suggest an algorithm and circuit that could be resolved as a compositional decision to proceed to one chord or another.

There are three distinct pitch class (PC) sets that form diminished seventh chords: {0,3,6,9}, {1,4,7,10}, and {2,5,8,11}. Any transposition of any of these results in a version of one of them. So, if a composer or improviser decides to use a diminished seventh chord, there are effectively only three PC sets to choose from. In practice, such a choice is not made arbitrarily. It is influenced by considerations of voice leading, short- or long-term harmonic targets (goals for resolution), and the degree of ambiguity desired. Some choices might be more stylistically appropriate or common than others, and the choice might well even be influenced by a related decision made elsewhere in the piece. Given the chord's vagrant ability to resolve to so many possible subsequent chords, however, it's literally impossible for a composer/improviser to make a "bad" choice, since a "good" subsequent chord exists in every case. So, our hypothetical composer/improviser (i.e., our computer program) need only choose one of the three options.

Once a PC set has been chosen, whether or not that choice was made with a particular root note in mind, any of the four pitch classes can be treated as the root (leading tone), which determines the chord's conceptual spelling and thus the likely root of its successor chord. For example, the PC set {0,3,6,9} can be spelled CE♭G♭B♭♭ (or B♯D♯F♯A), D♯F♯AC, F♯ACE♭, or ACE♭G♭ (or even G𝄪B♯D♯F♯), and thus can imply resolution to PC 1, 4, 7, or 10. Again, such choices are not made arbitrarily in actual practice. But the fact remains that, once the diminished seventh chord has been stated or implied, any of four pitch classes may be considered its root. The relative likelihood(s) of the subsequent chord(s) for the listener will again depend on melodic motion, a sense of the current key, stylistic traits of the music, and comparable occurrences elsewhere in the piece. Yet, at the local chord-to-chord level, each note may be the root with equal probability. So here, too, the composer/improviser, having chosen a pitch class set, can then confidently assign any of those four pitch classes to be the chord root.

---

[6] C.f., [6, 7, 8].

The leading tone diminished seventh chord, when not being used as a secondary or vagrant chord, can occur in a minor key by raising the subtonic note of the scale to be the leading tone, or it can occur in a major key by lowering the submediant note of the scale. Thus, a leading tone diminished seventh chord progresses quite readily to either a major or minor triad a semitone higher. This gives the computer composer one additional choice, in this case a binary one, to resolve to either a major or minor triad.

The preceding description outlines one very simple algorithm for chord progression, based on the ambiguity inherent in a particular chord type, and rooted in the ambiguities of chord spelling enabled by twelve-tone equal temperament. One chooses one of three pitch class sets, one then ascribes the role of chord root to one of the four pitch classes in that set, and one then chooses one of two chord qualities to which to resolve. This simple decision-making process provides twenty-four unique two-chord progressions (3x4x2=24), proceeding from any starting chord. The choices may be made probabilistically, either using equal probabilities with no particular consideration of directionality or key center, or using weighted probabilities to make some results more likely than others. The procedure can be carried out independently of considerations of voice leading, which may be determined by a separate algorithm.[7] In our example here, we'll use equal probability distribution.

The conventional way to choose one of *n* possibilities, each with equal likelihood, is to generate a random number within a range that's divisible into *n* equal regions. E.g. (in the C programming language):

```
int n = rand()%24;
```

where values 0 to 23 represent the 24 possible two-chord progressions. However, to demonstrate the applicability of quantum computing to this problem, we will use a method that requires only a very small number of binary (0-or-1) decisions. In quantum computing terms, we need only to measure one or two qubits for each decision.

---

[7] This is another deliberate oversimplification for the sake of this simple demonstration. In practice, a human composer assuredly considers voice-leading concurrently with harmonic progression, and chord choices may depend upon melodic considerations. In a future article we will demonstrate computer decision making based on voice-leading and melodic implications.

A choice between two possibilities, such as whether the final chord should be major or minor, is made by measuring a single qubit.[8] A choice among four possible alternatives, such as which note of the pitch class set will be considered the root of the chord, requires two qubits (figure 6).

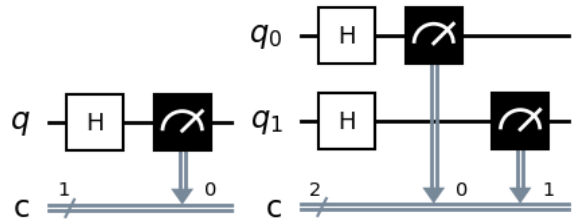

**Fig. 6.** One qubit $q$ and two qubits $q_0$ and $q_1$, with Hadamard gates H, measured as classical bits $c$.

In classical computing, a binary choice can be made by isolating a single bit of a random number; a four-way choice requires two bits, and so on for all powers of 2. For a choice among three options, such as deciding which pitch class set to use for the choice of diminished seventh chord, conventional computing does not provide a perfect way to make an equal-probability three-way choice using equal-probability binary choices, because 3 and 2 are mutually prime numbers. The most efficient substitute is to make a four-way choice with two random bits, and consider one of the four results to be a null choice that requires redoing the process [9]. This method theoretically has no upward bound, but in practice will require an average of $2.\bar{6}$ binary choices. This is where quantum computing provides a different method.

In quantum computing, the basis states of |0⟩ and |1⟩ can have unequal probabilities of being measured. This can be transformed by rotating the state vector on the Bloch sphere representation[9] of the qubit (figure 7).

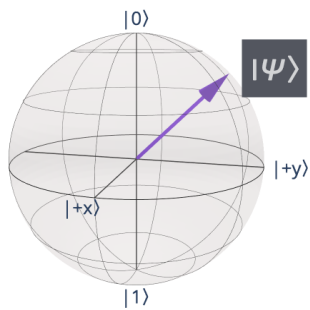

**Fig. 7.** Bloch sphere representation displaying an arbitrary state vector |Ψ⟩.

---

[8] This is achieved by use of a Hadamard gate (notated as H In a circuit diagram)—a quantum logical gate that puts the qubit in a state of superposition. Upon measurement, the state is collapsed into a classical state, either 0 or 1.

[9] The "Bloch sphere" is an abstract 3-dimensional representation of the computational space provided by a single qubit. C.f., [10].

In our specific case, rotating the qubit around the *x* axis in the amount of *arccos(⅓)* ($\cong$1.23) will give it a probability distribution of ⅔ |0⟩ and ⅓ |1⟩. When that qubit, $q_0$, is measured as 1, which happens ⅓ of the time, we can return that choice immediately as 01; otherwise, ⅔ of the time, we measure a second qubit, $q_1$, which has an equal probability distribution so that we return either 00 or 10 (figure 8).

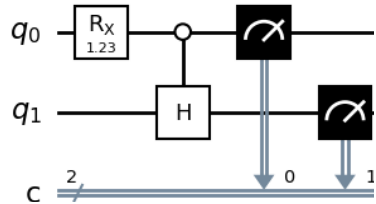

**Fig. 8.** Three-way choice with equal probability.

For implementing these experiments, we have been using the Quantum-computing Assisted Composition package (The QAC Toolkit) [11] in the Max programming environment. In that package, the **och.microqiskit** object provides a subset of the IBM open-source Quantum Information Science Kit (Qiskit) framework for quantum computing simulation. Specifying a circuit with the characteristics described above results in the three options 00, 01, and 10 being produced with equal probability (figure 9).

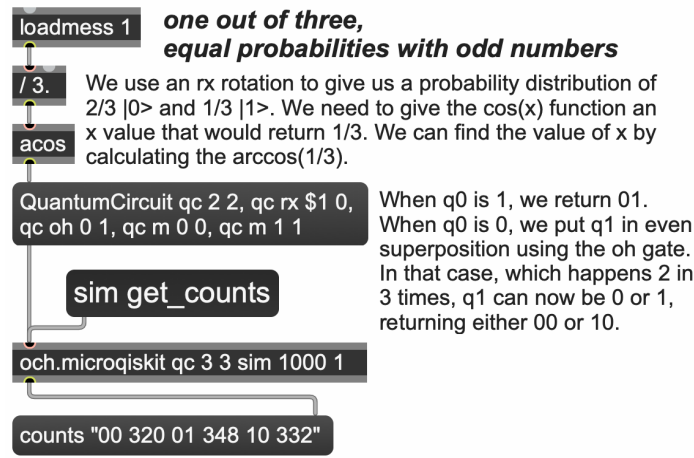

**Fig. 9.** Three-way choice with equal probability,
using rotation by *arccos(1/3)* (Rx) and anti-controlled (open) Hadamard gate (H).

Equipped with these quantum-computing-inspired methods for handling ambiguity, we can generate compositional choices of diminished seventh chords and their resolutions. We'll provide here two examples of a series of such choices made with this method, in conjunction with some very simple rules for voice leading (favoring conjunct motion, etc.). Starting with a C major triad, we use the computer to choose a subsequent diminished seventh chord, to which we attribute the status of chord root to one the four pitch classes (or, equivalently, in our program we attribute the role of root, third, fifth, or seventh to the lowest voice, which then determines the role that must be fulfilled by the other voices), and then resolve that chord to either a major or minor triad whose root is a semitone above. In these examples, every decision is made locally, which is to say without including consideration of any other factors such as larger harmonic context, long term goals, melodic contours, etc. The result is simply an alternating succession of diminished seventh chords and their consequent triad resolutions.

The first example (figure 10) shows a computer-composed progression consisting alternately of triads and diminished seventh chords. (Such frequent and consistent use of this alternating pattern would surely grow tiresome if carried on for too long, but it serves well here to demonstrate the idea and the methodology.) First, the program selected the PC set {0,3,6,9} using the above-described three-way decision method. It then attributed the role of seventh of the chord to the bass. The bass voice was already on a note of the selected PC set, so by staying where it is it optimally fulfills the criteria for conjunct voice leading, making C the seventh of the chord, thus making D♯ the chord root. This decision having been made, the program chose the most conjunct available voice leading to arrive at all the other needed pitches of the diminished seventh chord. It then decided that it should resolve to an E minor chord (selecting that over an E major chord), and resolved the voices "correctly": the leading tone resolves upward by semitone, the seventh of the chord resolves downward by semitone, the fifth of the chord by preference resolves downward by step to the third of the next chord, and the third of the chord in this case opted for the nearest neighbor as its tone of resolution. From that point, the entire process was repeated twice more to produce the four-measure example. Note that in this instance the entire progression can be analyzed as staying in C major.

**Fig. 10.** Example of computer decision making for chord progression.

Another run of the program using the same methodology takes us immediately to a very different destination (figure 11). In this case, the program selected the PC set {1,4,7,10}, and it gave the bass the role of the third of the chord. The most conjunct voice leading possible for the bass is to move from C to C♯ (from PC 0 to PC 1). For the listener, this will most logically imply ♯vii°7/ii, but because the computer has decided that C♯ is to be considered the third of the chord, the actual root of the chord is A♯, leading to B minor (or B major). Thus, already on the downbeat of measure 2, we land on a chord that does not belong to C major, indicating that we may be in the process of modulating. Subsequent computer decisions take us even further afield, to the far-related key of F♯ major.

**Fig. 11.** The same method produces different results.

## 4   Future Directions

For this experiment and this article, we have focused upon the diminished seventh chord as the most studied example of chord ambiguity. However, there is no dearth of other chord types that lend themselves well to this interpretation of being in a state of superposition. For example, the major-minor seventh chord type (a.k.a. the dominant seventh chord), which differs from the diminished seventh chord only by a single semitone, similarly has many possible interpretations and usages. It can be used as a dominant seventh chord resolving either authentically or deceptively in either a major or minor key, as a German augmented sixth chord in either major or minor, as a subtonic seventh chord in a minor key, as a subtonic "backdoor" dominant chord common in jazz and popular music either in major or in minor, as a tritone substitution for the dominant seventh chord (♭II7), as the result of double chromatic passing tones toward another dominant seventh chord, and so on.

Indeed, ambiguity of chord function is a crucial part of classical harmonic theory. Any major or minor triad belongs to more than one key, and therefore can be used to pivot from one key to another, a technique well known in classical music. The ambiguous harmonic function of such a pivot chord is revealed only upon its resolution to an unexpected chord or to a chord that does not belong to the initial key. Similarly, any diad can potentially imply membership in a variety of chords consisting of three or more pitches. (For example, the interval of a minor third can be the root and third of a minor triad, or the third and fifth of a major triad, or the fifth and seventh of a dominant seventh chord, or any two pitches of a diminished seventh chord, etc.) At the extreme limit, one might even say that every single pitch exists in a state of superposition, because a composer or improviser has free will to proceed from that pitch to anywhere at all. The notion that all musical sounds are in a state of multiplicity until they are "measured" by their resolution may be such a truism as to be almost useless, but it definitely alludes to the relationship between music theory and quantum theory, and can be used as a compositional tool, and as a mechanism for computer-generated music.

This idea of quantum compositional decision making has been extended to other chord types, into the realm of voice leading decisions and melodic composition, and non-standard theoretical models of harmony. We are undertaking an in-depth exploration of modeling the musical phenomenon of multiplicity of meaning and ambiguity as *superposition*. As a result of that correlation, we are evaluating the role of other key quantum concepts such as *entanglement* for future experimentation and more sophisticated computer-generated composition, creating new tools and methods for using quantum theory in a practical way. The role of equal temperament—equal divisions of the octave—as a catalyst for the functional ambiguity of pitch intervals implies future research in other equal-tempered divisions of the octave. Perhaps most importantly from the standpoint of creating new musical possibilities, the use of purely numerical music theory, dispensing with traditional note names, will enable this experimentation to go well beyond classical and jazz theory into other chord constructions and other harmonic orders.

Audio and software examples relevant to this article are available online at:
*https://harmony.quantumland.art/mcm24*

**Acknowledgments.** The work by Omar Costa Hamido is supported by the European Project IIMPAQCT, under grant agreement Nr. 101109258.

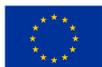

**Disclosure of Interests.** The authors have no competing interests to declare that are relevant to the content of this article.

# References

1. Bernhardt, C.: Quantum Computing for Everyone. The MIT Press, Cambridge MA (2019)
2. Larson, S. Musical Forces: Motion, Metaphor, and Meaning in Music. Indiana University Press, Bloomington IN (2012)
3. Meyer, L.: Meaning in Music and Information Theory. The Journal of Aesthetics and Art Criticism **15**(4), 412-424 (1957)
4. Stravinsky, I.: Poetics of Music in the Form of Six Lessons. Harvard University Press, Cambridge MA (1942)
5. Schoenberg, A.: Theory of Harmony (1911). Roy E. Carter (transl.) University of California Press, Los Angeles (1983)
6. Blutner, R.: Modelling tonal attraction: tonal hierarchies, interval cycles, and quantum probabilities. Soft Comput **21**, 1401–1419 (2017)
7. Souma, S.: Exploring the Application of Gate-Type Quantum Computational Algorithm for Music Creation and Performance. In: Miranda, E.R. (ed.) Quantum Computer Music. Springer, Cham (2022)
8. Fugiel, B. Quantum-like melody perception. Journal of Mathematics and Music, **17**:2, 319-331 (2023)
9. Knuth, D.E., Yao, A.C.: The complexity of nonuniform random number generation. In: Traub, J.F. (ed) Algorithms and Complexity: New Directions and Recent Results. Academic Press, New York (1976)

10. Nielsen, M.A., Chuang, I.L.: Quantum Computation and Quantum Information (2000). Cambridge University Press, Cambridge (2011)
11. Hamido, O.C.: QAC: Quantum-Computing Aided Composition. In: Miranda, E.R. (ed.) Quantum Computer Music. Springer, Cham (2022)